\documentclass[sigconf,screen]{acmart}
\AtBeginDocument{%
  }

\usepackage{xspace}
\usepackage{tikz,pifont,pgfplots}
\usepackage{xcolor, colortbl}
\usepackage{hyperref}
\usepackage[many]{tcolorbox}
\usepackage{graphicx}%
\usepackage{adjustbox}
\usepackage{mathrsfs}%
\usepackage[title]{appendix}%
\usepackage{xcolor}%
\usepackage{textcomp}%
\usepackage{manyfoot}%
\usepackage{booktabs}%
\usepackage{algorithm}%
\usepackage{algorithmicx}%
\usepackage{algpseudocode}%
\usepackage{listings}%
\usepackage{soul}
\usepackage{subcaption}
\usetikzlibrary{positioning}
\pgfplotsset{compat=1.18}

\newcommand*{\ie}{i.e.,\@\xspace}
\newcommand*{\eg}{e.g.,\@\xspace}


\newcommand*\circled[1]{\tikz[baseline=(char.base)]{\color{black} 
    \node[shape=circle,draw=cyan,fill=black!10!white,inner sep=.3pt] (char) {{{\texttt\textbf #1}}};}}

\newcommand*\circledw[1]{\tikz[baseline=(char.base)]{\color{black} 
    \node[shape=circle,draw=black,inner sep=.3pt] (char) {{{\texttt\textbf #1}}};}}

   \lstdefinestyle{searchstringstyle}{
	basicstyle=\ttfamily\scriptsize,
	captionpos=b,                    
	numbers=none,                    
	numbersep=4pt,                  
	showspaces=false,                
	showstringspaces=false,
	showtabs=false,                  
	tabsize=2,
	frame=single
}

\newcommand{\FairLMAck}{''FairLM: Addressing and Improving the Intersectional Bias of Large Language Models'' a project founded by the University of L'Aquila, 2025\xspace}

\setcopyright{acmlicensed}
\copyrightyear{2025}
\acmYear{2025}
\acmDOI{XXXXXXX.XXXXXXX}
\acmConference[Conference acronym 'XX]{Make sure to enter the correct
  conference title from your rights confirmation emai}{June 03--05,
  2018}{Woodstock, NY}

\newcommand{\our}{MANILA\xspace}
\newcommand{\fairworkflow}{fairness benchmarking workflow\xspace}

\usepackage{balance}

\begin{document}

\title{MANILA: A Low-Code Application to Benchmark Machine Learning Models and Fairness-Enhancing Methods}

\author{Giordano d'Aloisio}
\email{giordano.daloisio@univaq.it}
\orcid{0000-0001-7388-890X}
\affiliation{%
  \institution{University of L'Aquila}
  \city{L'Aquila}
  \country{Italy}
}

\renewcommand{\shortauthors}{d'Aloisio, G.}

\begin{abstract}
This paper presents MANILA, a web-based low-code application to benchmark machine learning models and fairness-enhancing methods and select the one achieving the best fairness and effectiveness trade-off. It is grounded on an Extended Feature Model that models a general fairness benchmarking workflow as a Software Product Line. The constraints defined among the features guide users in creating experiments that do not lead to execution errors. We describe the architecture and implementation of MANILA and evaluate it in terms of expressiveness and correctness. 
\end{abstract}

\ccsdesc[500]{Software and its engineering~Extra-functional properties}
\ccsdesc[300]{Computing methodologies~Machine learning}
\ccsdesc[300]{Software and its engineering~Designing software}
\keywords{Fairness, Machine Learning, Low Code, Benchmarking}


\acmBooktitle{Companion Proceedings of the 33rd ACM Symposium on the Foundations of Software Engineering (FSE '25), June 23--27, 2025, Trondheim, Norway}

\maketitle

\section{Introduction}
Machine learning (ML) systems are extensively employed in many domains nowadays. If we consider the impact that those applications have in our lives, it is clear how ensuring that those systems are of \emph{high} quality is of paramount importance. However, the quality of ML systems is not limited to their effectiveness (\ie prediction correctness) but also to other new quality attributes  \cite{muccini_software_2021,martinez-fernandez_software_2022,bosch_engineering_2021}. Among those, \textit{fairness} (\ie the absence of prejudice or discrimination of a system towards individuals or groups identified by a set of sensitive features \cite{mehrabi_survey_2021}) emerged as one of the most critical quality attributes of ML systems as also highlighted by some of the 17 Sustainable Development Goals (SDG) of the United Nations \cite{onu_sdg} like SDG 5 (Gender Equality) or SDG 10 (Reduced Inequalities).

In recent years, many low-code methods have been developed to automate some phases of ML system development and help less technical users, \eg \cite{RONKKO201513,6086455,HE2021106622}. However, a low-code approach that also considers the model's fairness and guides less technical users in the benchmark of different ML models and fairness-enhancing methods is still missing. Indeed, there is a need for low-code approaches to help democratize the quality-based development of ML systems \cite{lee_landscape_2021,nguyen_literature_2024}

With this paper, we aim to fill this gap by proposing MANILA, a web-based low-code application to benchmark ML models and fairness-enhancing methods and select the combination that achieves the best fairness-effectiveness trade-off. The theoretical foundation of MANILA is an Extended Feature Model (ExtFM) that models a general fairness benchmarking workflow as a Software Product Line (SPL). The constraints among the features in the ExtFM have been reported in the web application and guide users in the definition of experiments that are always executable - \ie they do not lead to execution errors. For space constraints, we leave details on the theoretical foundations of MANILA and to the ExtFM to our previous work \cite{manila2023}.

MANILA can be employed by practitioners and researchers who have knowledge of ML but lack expertise in the fairness domain and want to benchmark different combinations of ML models and fairness-enhancing methods to identify the best one.

\our is publicly available in the SoBigData research infrastructure \cite{grossi_data_2018}\footnote{\url{https://sobigdata.d4science.org/group/sobigdata.it/manila-univaq} (registration needed)}, and its source code is available on GitHub \cite{manila-app}. A documentation site presenting a hands-on tutorial on MANILA is also publicly available \cite{manila-doc}. A video demonstration of \our is available online \cite{manila_video}.

\section{MANILA}\label{sec:approach}
\our is a low-code application to benchmark ML models and fairness-enhancing methods and to select the one that offers the best effectiveness and fairness trade-off. As outlined in \cite{manila2023}, each \fairworkflow comprises a set of features acting as \textit{variation points}, differentiating them from one another. For this reason, we can think of this family of experiments as a Software Product Line (SPL) specified by an Extended Feature Model \cite{manila2023}. 

\begin{figure}[ht!]
    \centering
    \includegraphics[width=\linewidth]{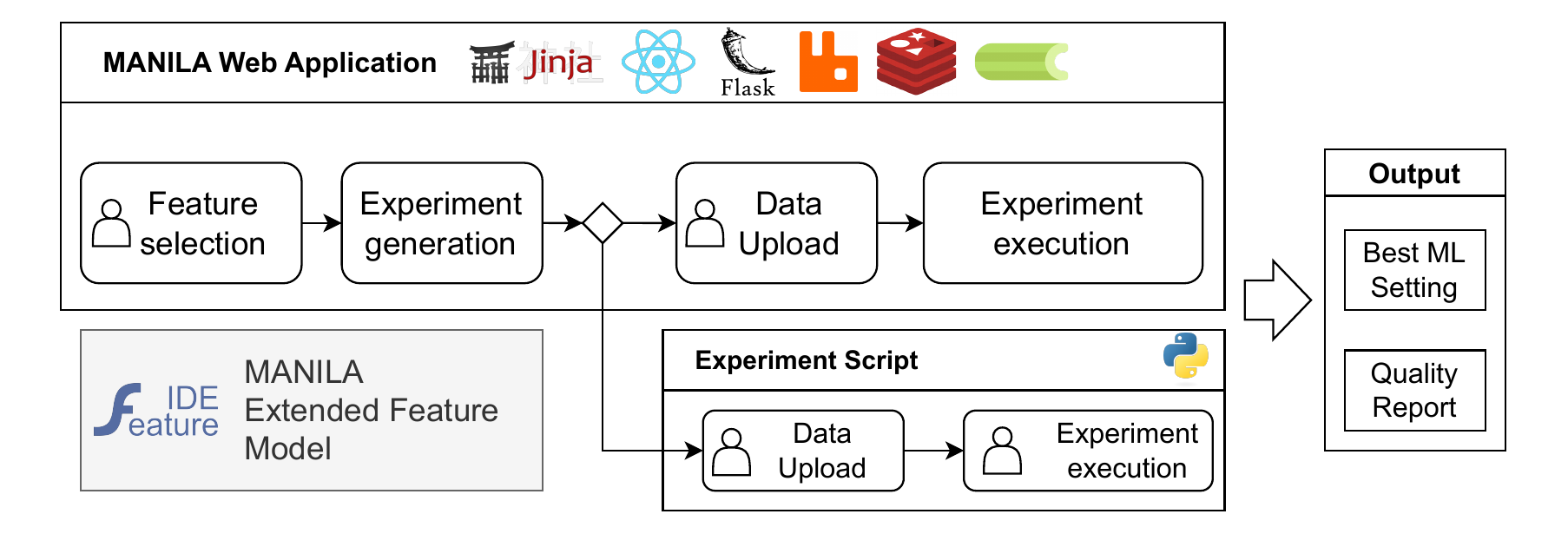}
    \caption{\our high-level overview}
    \label{fig:manila}
\end{figure}

Figure \ref{fig:manila} details a high-level picture of \our, where each rounded box represents a step in the \fairworkflow, while square boxes represent artifacts. \our has been implemented as a low-code web-based application through which all the steps of the \fairworkflow can be performed. Near each artifact, we report the technologies involved in its implementation.

The first step in the development process is feature selection, in which the user selects all the components of the \fairworkflow through a dedicated web form. Next, a Python script implementing the benchmarking experiment is automatically generated from the selected features. The constraints defined among features guide the user in defining experiments that are always executable (i.e., do not lead to execution errors) \cite{manila2023}.
The experiment performs a grid search \cite{lerman1980fitting} over the selected combinations of ML models and fairness-enhancing methods and computes the selected fairness and effectiveness metrics. The experiment can be downloaded for local execution or directly executed on the server. After its execution, it returns: \textit{i)} a quality report reporting for each fairness-enhancing method and ML model the related metrics; \textit{ii)} the best combination identified using the selected trade-off strategy.

As described in \cite{manila2023}, the foundation of \our is an Extended Feature Model (ExtFM) that models the \fairworkflow as an SPL. The ExtFM is a template of all possible experiments a user can perform and guides them through selecting features comprising a \fairworkflow. We used the ExtFM as a formalism to reason about the different relationships and constraints among features before implementing them in the web application. 

The web application has been implemented using the React Javascript library \cite{react} for the frontend and the Flask Python library \cite{flask} for the backend. Moreover, we employ the Celery worker engine \cite{celery}, with the RabbitMQ message broker \cite{rabbitmq} and the Redis database \cite{redis}, to run the experiments asynchronously on the server. This way, we avoid overloading the server in case of multiple experiment runs.


\subsection{Feature Selection}\label{sec:feature_sel}

\begin{figure}[tb]
    \centering
    \includegraphics[width=\linewidth]{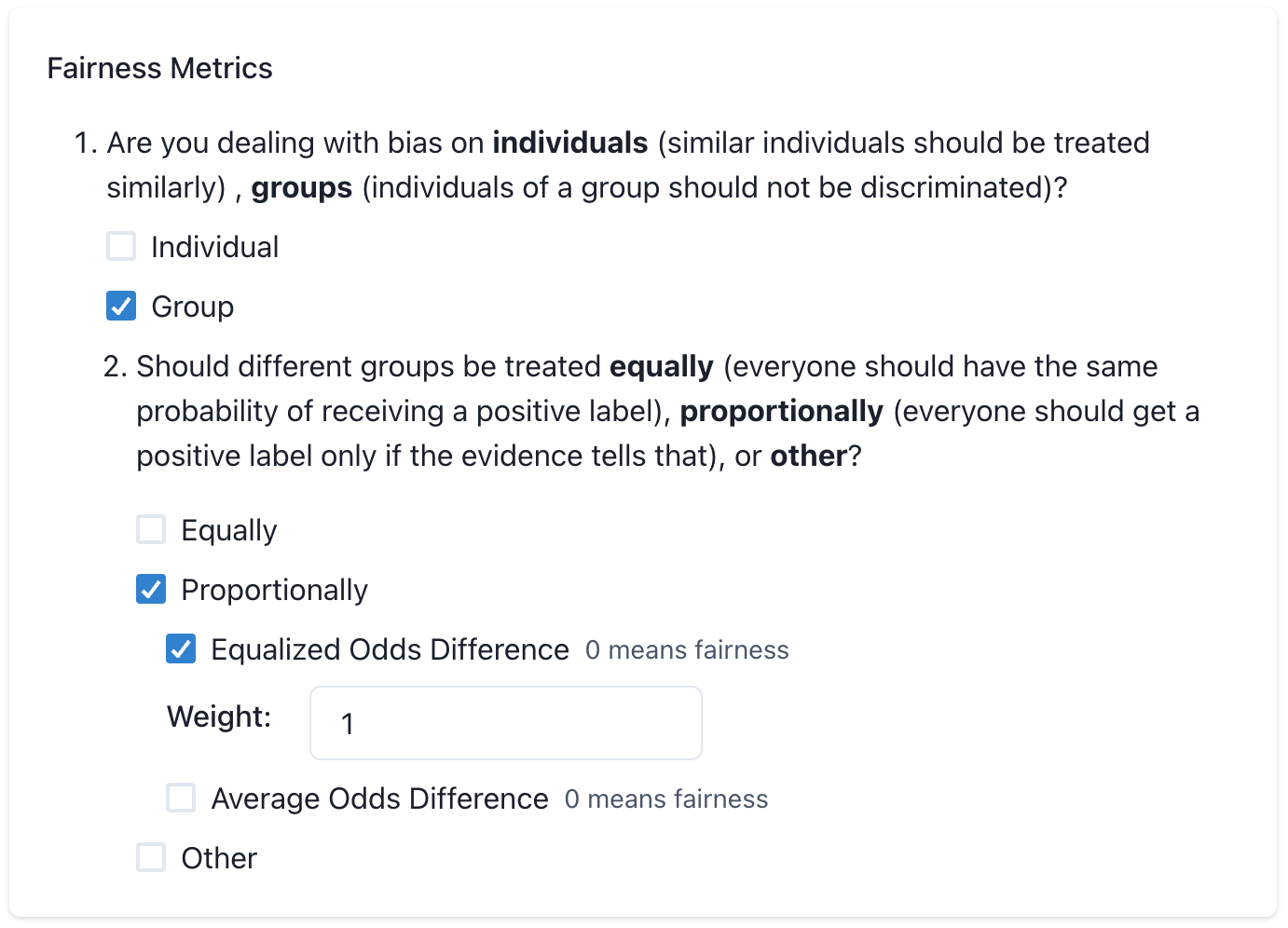}
    \caption{Fairness metric selection}
    \label{fig:metric-selection}
\end{figure}

The feature selection step has been implemented in \our through a web form that includes all the features and the constraints defined in the ExtFM.
The form consists of seven subsections, each corresponding to one of the macro features defined in the ExtFM. Each subsection includes all the concrete (i.e., non-abstract \cite{apel2016feature}) child features related to the respective macro feature in the ExtFM. The sections in the form are the following. \textbf{\circled{1} Dataset:} This section contains features to specify information about the dataset (e.g., the label or sensitive features). \textbf{\circled{2} Data Scaler:} This section includes all data scaler algorithms from the \textit{scikit-learn} library \cite{scikit-learn} to pre-process data before training a model. \textbf{\circled{3} ML Model:} This section allows the users to select the ML models to benchmark. All the ML models for classification and regression from the \textit{scikit-learn} library are listed. \textbf{\circled{4} Fairness Methods:} This section allows the selection of the fairness-enhancing methods to benchmark. We included all models from \texttt{aif360} \cite{bellamy_ai_2019} and \texttt{Fairlearn} \cite{bird2020fairlearn} libraries and the \textit{Debiaser for Multiple Variables} algorithm \cite{daloisio_debiaser_2023}. \textbf{\circled{5} Metrics:} This section includes all metrics that can be employed to evaluate the effectiveness and fairness of a given ML model and fairness-enhancing method combination. Concerning effectiveness metrics, we included all the classification and regression metrics available in the \texttt{scikit-learn} library. For fairness assessment, we included all the metrics from the \texttt{aif360} library. In addition, given the magnitude of fairness definitions and metrics available \cite{mehrabi_survey_2021}, following \cite{10.1145/3550355.3552401}, we included a set of questions to help the data scientist select the fairness metrics more suited for their use case. The questions are reported in Figure \ref{fig:metric-selection}. \textbf{\circled{6} Tradeoff Strategy:} This section lists the strategies that can be adopted to identify the best ML model and fairness-enhancing method combination. In particular, the best setting can be identified by MANILA using an aggregation function of the selected metrics (e.g., mean, weighted sum, or harmonic mean) or through a Pareto-front of the best solutions \cite{10.1162/EVCO_a_00128}. \textbf{\circled{7} Validation:} This last section allows the selection of strategies for cross-validation of the different ML models and fairness-enhancing methods combination.

Children with an \textit{alternative} relationship in the ExtFM \cite{apel2016feature} are implemented in the form either through a radio group or by a logical condition among fields. In all other cases, features in the ExtFM have been implemented as checkbox fields in the form. Additional attributes related to the features (like the \textit{Label Name} or \textit{Positive Value} fields) have been implemented as text fields, which may be mandatory or not, depending on the case. 
\begin{figure}[ht!]
    \centering
    \includegraphics[width=.85\linewidth]{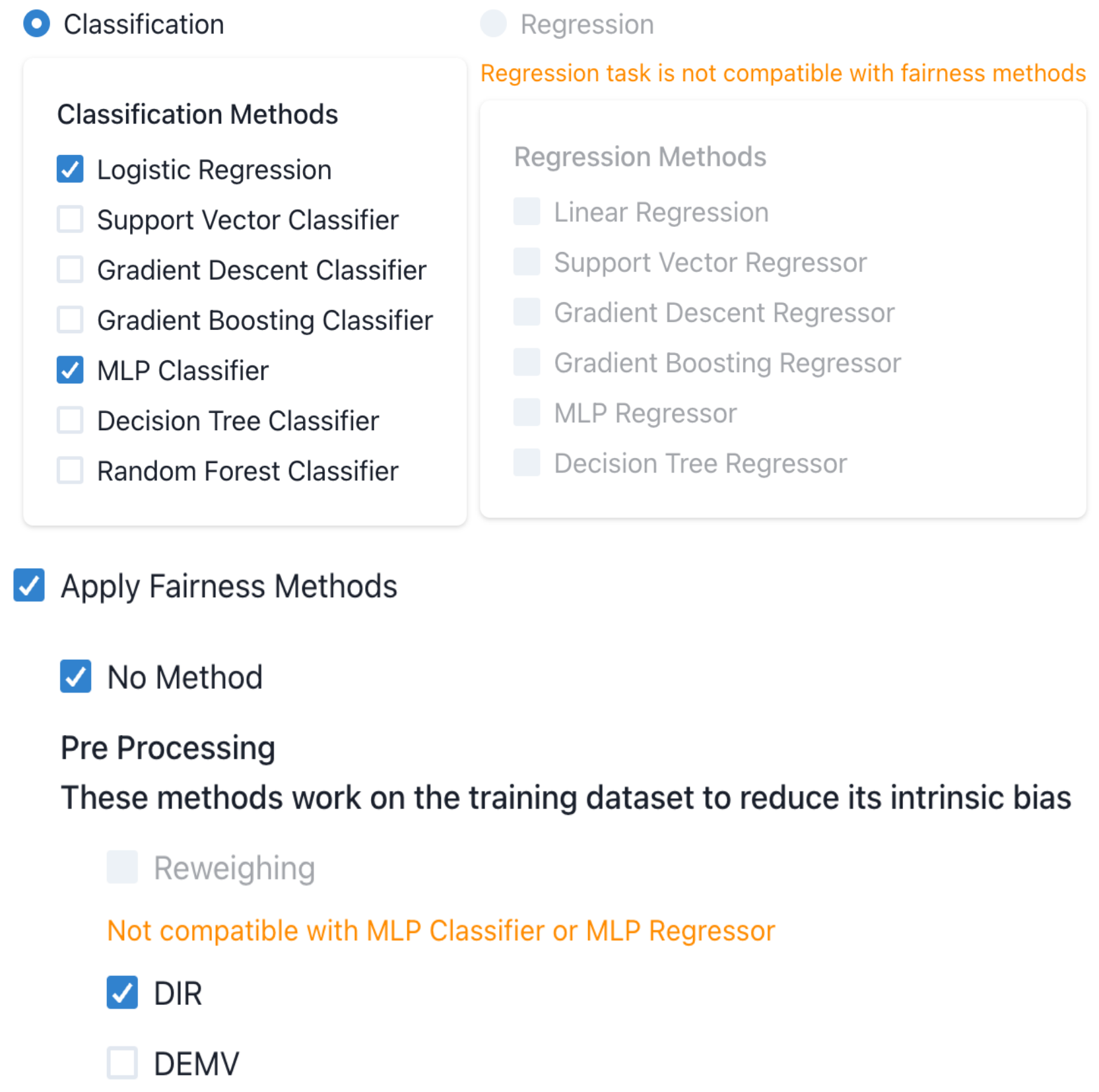}
    \caption{Example of web form cross-tree constraints}
    \label{fig:web-constr}
\end{figure}
Additionally, the cross-tree constraints defined in the ExtFM have been implemented as logical constraints among the different form fields. Figure \ref{fig:web-constr} shows an example of such constraints. In the figure, it can be seen how the \textit{Regression} field has been disabled. This is due to two reasons: first, the \textit{Classification} ML task has already been selected, and second, the \textit{Regression} task is incompatible with the fairness methods included so far. Hence, since fairness-enhancing methods have already been selected in another section of the form, ML methods for regression can not be selected. Otherwise, they would lead to a non-executable experiment. This constraint is also shown to the user through a message reporting that "\textit{Regression task is incompatible with fairness methods}." In addition, note how the \textit{Reweighing} fairness method has been disabled as well. This is because the \textit{Reweighing} method is not compatible with the \textit{MLP Classifier} ML method that has already been selected. This constraint is also reported to the user, reporting that Reweighing is "\textit{not compatible with MLP Classifier or MLP Regressor}."

Finally, the form enables users to upload their dataset for running the generated experiment on the server. Below this field, two buttons are available: one for downloading the generated code and the other for executing the experiment on the server. 

\subsection{Experiment Generation and Execution}\label{sec:code_gen}

After selecting a set of features that meet all the form's constraints, the experiment can be executed. As shown in Figure \ref{fig:manila}, the experiment can be either downloaded for local execution or run directly on the server. The experiment performs a grid search across all combinations of the selected ML models and fairness-enhancing methods, computing the selected metrics for each combination. It is worth noticing how the space of the search (\ie the possible combinations of ML models and fairness-enhancing methods) is always relatively small, given the constraints imposed by the ExtFM. 

If the user chooses to download the experiment, MANILA generates the corresponding Python implementation relying on the Jinja template engine \cite{jinja2}. Additionally, it generates the \texttt{environment.yml} file needed to create the \textit{conda} environment with all the required libraries \cite{miniconda}. The experiment can be executed locally by running the following command:  
\begin{lstlisting}[style=searchstringstyle]
$ python main.py -d <DATASET PATH>
\end{lstlisting}
Otherwise, it can be called through a REST API or any other interface such as a desktop application or a Scientific Workflow Management System like \textit{KNIME} \cite{liu2015survey,knime_2009}. This generality of our experimental workflow makes it very flexible and suitable for many use cases. After the execution, the code returns the quality report in CSV format. In addition, the best ML model (identified using the selected trade-off strategy) is trained with the full input dataset and by applying the best fairness-enhancing method. The ML model returned by the experiment is saved as a \texttt{dill} file \cite{pickle}. We have chosen this format since it is an extension of the standard \texttt{pickle} format that enables the storage of more complex objects. 

\begin{figure}[tb]
    \centering
    \includegraphics[width=\linewidth]{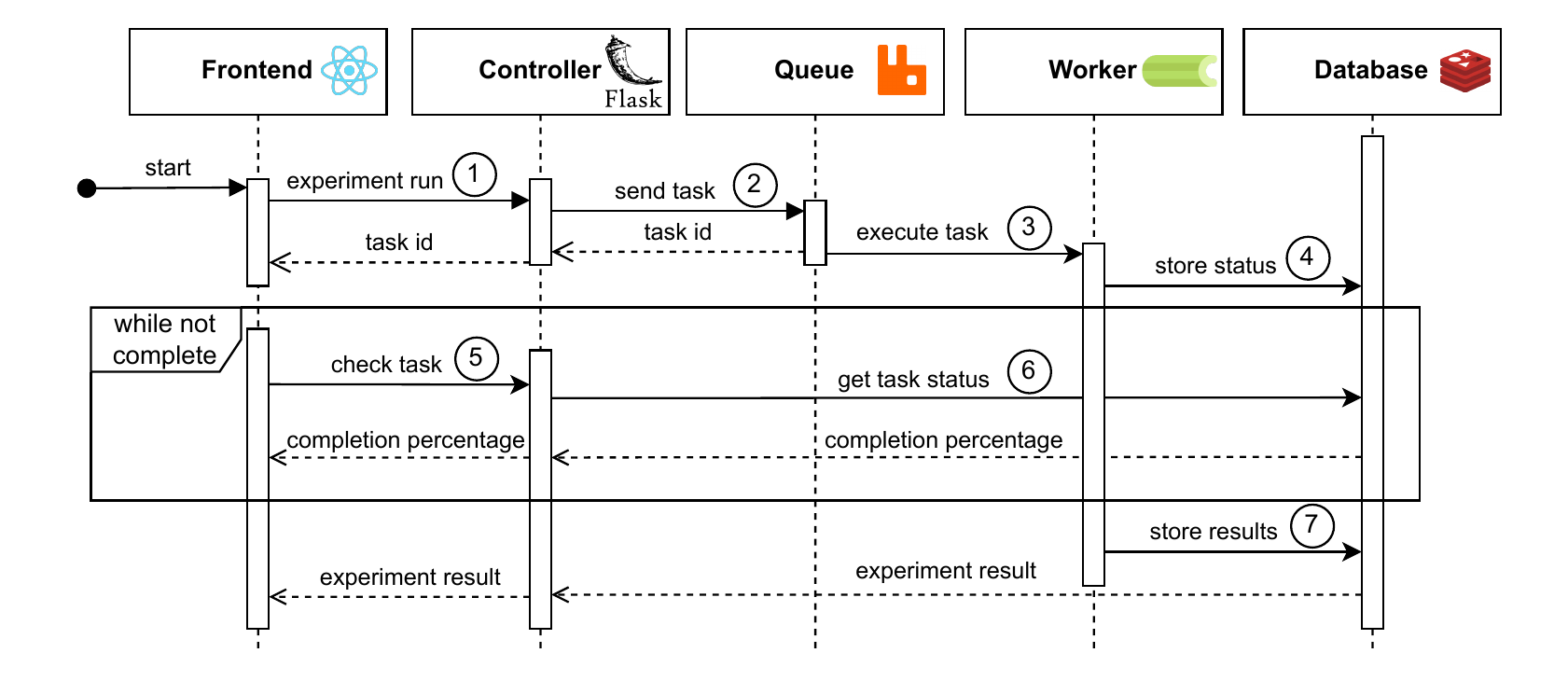}
    \caption{Online experiment execution}
    \label{fig:exp_exc}
\end{figure}

On the contrary, if the user chooses to run the experiment online, it is executed asynchronously using the Celery task queue system \cite{celery}. Figure \ref{fig:exp_exc} shows the online execution process. First, the frontend makes a call to the app controller through its REST API (step \circledw{1} in Figure \ref{fig:exp_exc}). The controller sends the task to a RabbitMQ queue and returns the task ID to the frontend (step \circledw{2} in Figure \ref{fig:exp_exc}). Next, the task is executed by a Celery worker, and its status is stored in a Redis database (steps \circledw{3} and \circledw{4} in Figure \ref{fig:exp_exc}). At the same time, until the experiment execution is not completed, the frontend periodically asks the controller about the task status (steps \circledw{5} and \circledw{6} in Figure \ref{fig:exp_exc}). If the experiment is still in progress, a completion percentage is returned. Otherwise, when the execution is completed, the Celery worker saves the results on Redis (step \circledw{7} in Figure \ref{fig:exp_exc}), and the results are returned to the frontend. The results are shown in a dedicated page. The result page shows the metrics in a bar chart and in a tabular way. Additionally, it allows the download of the fully trained best ML model in \texttt{dill} format and the computed metrics as a CSV file, respectively. 

\section{Evaluation}\label{sec:evaluation}
In this section, we describe the initial evaluation we performed on \our. 
Following previous work \cite{10.1145/3550355.3552401}, we evaluate \our in terms of \textit{expressiveness} and \textit{correctness}. In particular, we reproduce with \our the experimental evaluation described in \cite{daloisio_debiaser_2023} to evaluate the DEMV bias mitigation algorithm. More in detail, the \textit{expressiveness} is assessed by proving that \our can replicate the selected fairness evaluation. In contrast, \textit{correctness} is assessed by showing that the results obtained with the experiments generated by \our are comparable to the original ones.

We replicate three specific experiments performed on DEMV (\ie the ones shown in Tables 16, 18, and 19 of \cite{daloisio_debiaser_2023}). We have chosen to replicate these experiments among all the ones conducted in the paper because they provide the highest combination of ML methods, fairness-enhancing methods, and metrics.

The \textit{expressiveness} is evaluated by assessing if \our can properly reproduce the described experiments. The \textit{correctness} is evaluated by assessing if the results of the experiments generated by \our are close to the ones reported in the original experiments. More in detail, following previous work \cite{10.1145/3550355.3552401}, the results of the generated experiments should be within the standard deviation range of the results reported in Tables 16, 18, and 19, and there should not be a statistically significant difference between the results. The replication package of the experiments is available in our Github repository \cite{manila-app}.

Additionally, we plan to evaluate MANILA's \textit{usefulness} (\ie how much \our is perceived as useful) and \textit{usability} (\ie how much \our is easy to use) by performing a user evaluation involving people with diverse level of expertise in Software Engineering and Machine Learning-\ie computer science master students, researchers, and practitioners. Following standard guidelines \cite{runeson2012case}, we plan to ask each group of evaluators to perform two fairness benchmarking experiments: one using Python code and one using \our. Eventually, we will collect feedback on the positive and negative aspects of conducting the benchmarking process in Python and with \our. Moreover, we plan to compare the capabilities of \our against similar baseline approaches \cite{10.1145/3550355.3552401,johnson_fairkit-learn_2022}.

\paragraph{Expressiveness Evaluation}

We were able to correctly reproduce all the experiments. In particular, following the steps described in Section \ref{sec:approach}, we first specified the features of the experiments (\ie ML models, fairness-enhancing methods, and metrics) from the graphical interface of \our. Next, we downloaded the generated codes to execute them locally. Finally, we ran the experiments to obtain the results. In total, we generated and executed ten different experiments (\ie one for each input dataset).
Being a low-code platform, \our does not require to write any line of code to implement the given experiments. In contrast, the original experiments required almost 200 lines of code, as seen in the repository linked in the original paper \cite{daloisio_debiaser_2023}.

\paragraph{Correctness Evaluation}

Table \ref{tab:pvalues} reports, for each metric of each experiment, the \textit{p-values} of the non-parametric Kruskal-Wallis H test performed between the results of the original experiments and the ones obtained by executing the code generated by \our. 
\begin{table}[tb]
    \centering
   \caption{\textit{p-values} of the Kruskal-Wallis H test for each experiment}
    \label{tab:pvalues}
    \begin{tabular}{c|c|c|c|c|c|c}
        \toprule
         &  \textbf{SP} &   \textbf{AO} & \textbf{ZO Loss} &  \textbf{DI} &       \textbf{Acc} &     \textbf{H-Mean} \\
        \midrule
        \textbf{Exp. 1} & 0.84 & 0.08 & 0.84 & 0.45 & 0.38 & 0.53 \\\midrule
        \textbf{Exp. 2} & 0.10 & 0.71 & 0.31 & 0.43 & 0.27 & 0.71 \\\midrule
        \textbf{Exp. 3} & 1.00 & 0.57 & 0.16 & 0.96 & 0.72 & 0.75 \\\bottomrule
    \end{tabular}
\end{table}
From the table, it can be seen that all the \textit{p-values} are $> 0.05$, meaning that all the metrics obtained by running the code generated by \our are not statistically different from the original ones.
\begin{figure}[ht!]
    \centering
    \includegraphics[width=\linewidth]{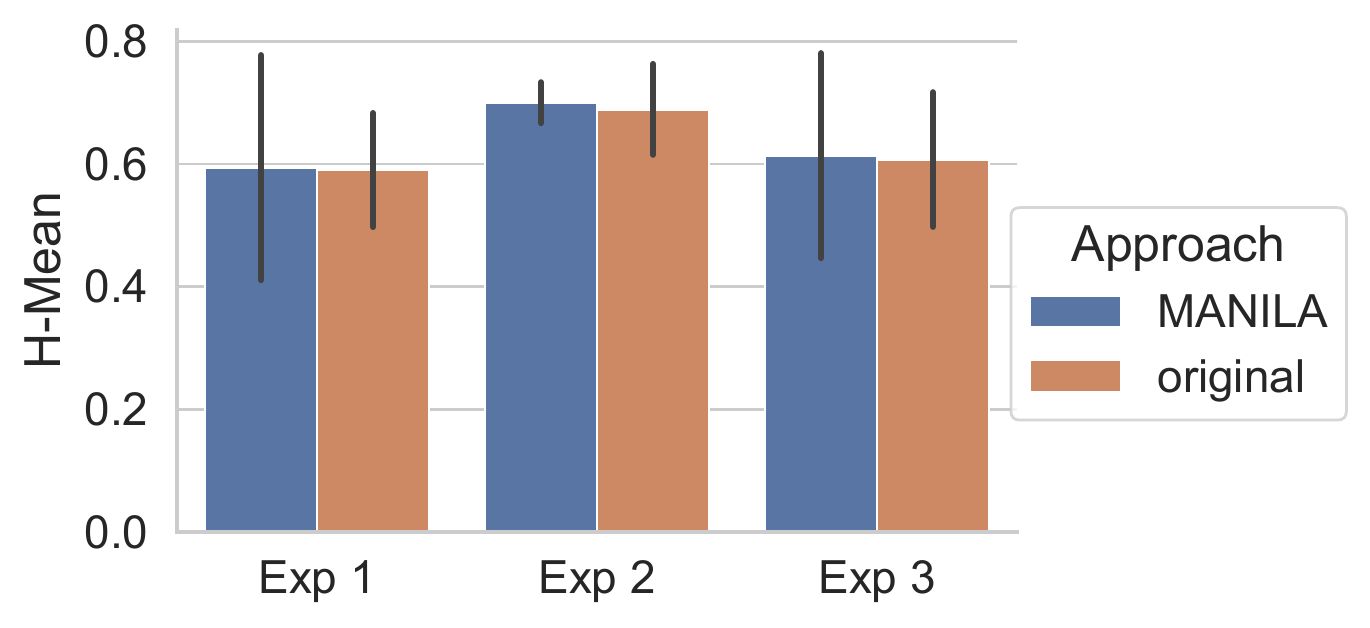}
    \caption{Aggregated H-Means of the original experiments and MANILA's ones}
    \label{fig:comparison}
\end{figure}
In addition, Figure \ref{fig:comparison} reports a comparison of the aggregated h-means \cite{Ferger193136} of the three experiments. As shown, on average, the results of the three experiments generated by \our are very close to the original ones and are within the standard deviation range.


\section{Related Work}\label{sec:related}
Low-code approaches like \textit{Aequitas} \cite{saleiro2018aequitas} Google's \textit{What-If Tool} \cite{wexler2019if}, Themis \cite{angell_themis_2018}, and MODNESS \cite{d2024fair} have been proposed to ease and democratize the fairness assessment of ML models. However, they do not allow to benchmark the combination of different ML models and fairness-enhancing methods. 

A similar search-based approach to identify the best combination of the ML classifier and the fairness enhancing method is the \textit{Fairkit-learn} Python library proposed by Johnson and Brun \cite{johnson_fairkit-learn_2022}. The library employs a grid-search approach to identify the optimal combination of fairness-enhancing methods and ML classifiers. In contrast, our work proposes a low-code application aimed at democratizing this process, making it more accessible for less technical users \cite{lee_landscape_2021}. Additionally, \our provides guidance for defining and generating executable experiments, ensuring they don't lead to execution errors. 

Yohannis and Kolovos propose a model-driven framework for bias assessment and mitigation \cite{10.1145/3550355.3552401}. The authors have first defined a meta-model to represent bias measurement and mitigation experiments. Then, they defined a Domain Specific Language (DSL) to specify such evaluations. The proposed tool automatically generates a Python script implementation from a script model defined using such a DSL. However, their approach does not guide the user in the definition of experiments that are always executable. By relying on the ExtFM formalism, \our fills this gap. Additionally, \our allows the online execution of the experiment. Thus, it does not require the user to execute the evaluation manually.

\section{Conclusion}
In this paper, we presented MANILA, a low-code web application that can be employed by practitioners and researchers with less knowledge of fairness to benchmark different ML models and fairness-enhancing method combinations. MANILA is grounded on the ExtFM formalism, which models a general fairness benchmarking workflow as an SPL where the variation points are the different components of the experiment, such as the ML models, fairness-enhancing methods, metrics, or trade-off strategies. The constraints defined among features guide users in creating always executable experiments. We described the system's architecture and evaluated it in terms of \textit{expressiveness} and \textit{correctness} by replicating a real experimental evaluation. 

\balance
\begin{acks}
    This work is partially supported by \FairLMAck.
\end{acks}

\bibliographystyle{ACM-Reference-Format}
\bibliography{bibliography}


\begin{thebibliography}{40}


\ifx \showCODEN    \undefined \def \showCODEN     #1{\unskip}     \fi
\ifx \showDOI      \undefined \def \showDOI       #1{#1}\fi
\ifx \showISBNx    \undefined \def \showISBNx     #1{\unskip}     \fi
\ifx \showISBNxiii \undefined \def \showISBNxiii  #1{\unskip}     \fi
\ifx \showISSN     \undefined \def \showISSN      #1{\unskip}     \fi
\ifx \showLCCN     \undefined \def \showLCCN      #1{\unskip}     \fi
\ifx \shownote     \undefined \def \shownote      #1{#1}          \fi
\ifx \showarticletitle \undefined \def \showarticletitle #1{#1}   \fi
\ifx \showURL      \undefined \def \showURL       {\relax}        \fi
\providecommand\bibfield[2]{#2}
\providecommand\bibinfo[2]{#2}
\providecommand\natexlab[1]{#1}
\providecommand\showeprint[2][]{arXiv:#2}

\bibitem[min({[n.\,d.]})]%
        {miniconda}
 \bibinfo{year}{[n.\,d.]}\natexlab{}.
\newblock \bibinfo{title}{Conda website}.
\newblock
\newblock
\urldef\tempurl%
\url{https://docs.conda.io/}
\showURL{%
\tempurl}


\bibitem[cel(2025)]%
        {celery}
 \bibinfo{year}{2025}\natexlab{}.
\newblock \bibinfo{title}{Celery Library}.
\newblock
\newblock
\urldef\tempurl%
\url{https://docs.celeryq.dev/en/stable}
\showURL{%
\tempurl}


\bibitem[pic(2025)]%
        {pickle}
 \bibinfo{year}{2025}\natexlab{}.
\newblock \bibinfo{title}{Dill documentation}.
\newblock
\newblock
\urldef\tempurl%
\url{https://dill.readthedocs.io/}
\showURL{%
\tempurl}


\bibitem[fla(2025)]%
        {flask}
 \bibinfo{year}{2025}\natexlab{}.
\newblock \bibinfo{title}{Flask Library}.
\newblock
\newblock
\urldef\tempurl%
\url{https://flask.palletsprojects.com/}
\showURL{%
\tempurl}


\bibitem[rab(2025)]%
        {rabbitmq}
 \bibinfo{year}{2025}\natexlab{}.
\newblock \bibinfo{title}{RabbitMQ Website}.
\newblock
\newblock
\urldef\tempurl%
\url{https://www.rabbitmq.com/}
\showURL{%
\tempurl}


\bibitem[rea(2025)]%
        {react}
 \bibinfo{year}{2025}\natexlab{}.
\newblock \bibinfo{title}{React Library}.
\newblock
\newblock
\urldef\tempurl%
\url{https://react.dev/}
\showURL{%
\tempurl}


\bibitem[red(2025)]%
        {redis}
 \bibinfo{year}{2025}\natexlab{}.
\newblock \bibinfo{title}{Redis Website}.
\newblock
\newblock
\urldef\tempurl%
\url{https://redis.io/}
\showURL{%
\tempurl}


\bibitem[Angell et~al\mbox{.}(2018)]%
        {angell_themis_2018}
\bibfield{author}{\bibinfo{person}{Rico Angell}, \bibinfo{person}{Brittany Johnson}, \bibinfo{person}{Yuriy Brun}, {et~al\mbox{.}}} \bibinfo{year}{2018}\natexlab{}.
\newblock \showarticletitle{Themis: automatically testing software for discrimination}. In \bibinfo{booktitle}{\emph{Proceedings of the 2018 26th {ACM} {Joint} {Meeting} on {European} {Software} {Engineering} {Conference} and {Symposium} on the {Foundations} of {Software} {Engineering}}} \emph{(\bibinfo{series}{{Esec}/{fse} 2018})}. \bibinfo{publisher}{Association for Computing Machinery}, \bibinfo{address}{New York, NY, USA}, \bibinfo{pages}{871--875}.
\newblock
\showISBNx{978-1-4503-5573-5}
\urldef\tempurl%
\url{https://doi.org/10.1145/3236024.3264590}
\showDOI{\tempurl}


\bibitem[Apel et~al\mbox{.}(2016)]%
        {apel2016feature}
\bibfield{author}{\bibinfo{person}{Sven Apel}, \bibinfo{person}{Don Batory}, \bibinfo{person}{Christian K{\"a}stner}, {et~al\mbox{.}}} \bibinfo{year}{2016}\natexlab{}.
\newblock \bibinfo{booktitle}{\emph{Feature-oriented software product lines}}.
\newblock \bibinfo{publisher}{Springer}.
\newblock


\bibitem[Bellamy et~al\mbox{.}(2019)]%
        {bellamy_ai_2019}
\bibfield{author}{\bibinfo{person}{R.~K.~E. Bellamy}, \bibinfo{person}{K. Dey}, \bibinfo{person}{M. Hind}, {et~al\mbox{.}}} \bibinfo{year}{2019}\natexlab{}.
\newblock \showarticletitle{{AI} {Fairness} 360: {An} extensible toolkit for detecting and mitigating algorithmic bias}.
\newblock \bibinfo{journal}{\emph{IBM Journal of Research and Development}} \bibinfo{volume}{63}, \bibinfo{number}{4/5} (\bibinfo{year}{2019}), \bibinfo{pages}{4:1--4:15}.
\newblock
\showISSN{0018-8646}
\urldef\tempurl%
\url{https://doi.org/10.1147/jrd.2019.2942287}
\showDOI{\tempurl}
\newblock
\shownote{Conference Name: IBM Journal of Research and Development}.


\bibitem[Berthold et~al\mbox{.}(2009)]%
        {knime_2009}
\bibfield{author}{\bibinfo{person}{Michael~R. Berthold}, \bibinfo{person}{Nicolas Cebron}, \bibinfo{person}{Fabian Dill}, {et~al\mbox{.}}} \bibinfo{year}{2009}\natexlab{}.
\newblock \showarticletitle{KNIME - the Konstanz Information Miner: Version 2.0 and Beyond}.
\newblock \bibinfo{journal}{\emph{SIGKDD Explor. Newsl.}} \bibinfo{volume}{11}, \bibinfo{number}{1} (\bibinfo{year}{2009}), \bibinfo{pages}{26–31}.
\newblock
\showISSN{1931-0145}
\urldef\tempurl%
\url{https://doi.org/10.1145/1656274.1656280}
\showDOI{\tempurl}


\bibitem[Bird et~al\mbox{.}(2020)]%
        {bird2020fairlearn}
\bibfield{author}{\bibinfo{person}{Sarah Bird}, \bibinfo{person}{Miro Dud{\'i}k}, \bibinfo{person}{Richard Edgar}, {et~al\mbox{.}}} \bibinfo{year}{2020}\natexlab{}.
\newblock \bibinfo{booktitle}{\emph{Fairlearn: A toolkit for assessing and improving fairness in {AI}}}.
\newblock \bibinfo{type}{{T}echnical {R}eport} Msr-tr-2020-32. \bibinfo{institution}{Microsoft}.
\newblock
\urldef\tempurl%
\url{https://www.microsoft.com/en-us/research/publication/fairlearn-a-toolkit-for-assessing-and-improving-fairness-in-ai/}
\showURL{%
\tempurl}


\bibitem[Bosch et~al\mbox{.}(2021)]%
        {bosch_engineering_2021}
\bibfield{author}{\bibinfo{person}{Jan Bosch}, \bibinfo{person}{Helena~Holmström Olsson}, {and} \bibinfo{person}{Ivica Crnkovic}.} \bibinfo{year}{2021}\natexlab{}.
\newblock \bibinfo{title}{Engineering {AI} {Systems}: {A} {Research} {Agenda}}.
\newblock
\newblock
\urldef\tempurl%
\url{https://doi.org/10.4018/978-1-7998-5101-1.ch001}
\showDOI{\tempurl}
\newblock
\shownote{ISBN: 9781799851011 Pages: 1-19 Publisher: IGI Global}.


\bibitem[d'Aloisio(2025a)]%
        {manila-doc}
\bibfield{author}{\bibinfo{person}{Giordano d'Aloisio}.} \bibinfo{year}{2025}\natexlab{a}.
\newblock \bibinfo{title}{{MANILA Documentation}}.
\newblock
\newblock
\urldef\tempurl%
\url{https://manila-fairness.github.io/}
\showURL{%
\tempurl}


\bibitem[d'Aloisio(2025b)]%
        {manila-app}
\bibfield{author}{\bibinfo{person}{Giordano d'Aloisio}.} \bibinfo{year}{2025}\natexlab{b}.
\newblock \bibinfo{title}{{MANILA Github Repository}}.
\newblock
\newblock
\urldef\tempurl%
\url{https://github.com/giordanoDaloisio/manila-web}
\showURL{%
\tempurl}


\bibitem[d'Aloisio(2025c)]%
        {manila_video}
\bibfield{author}{\bibinfo{person}{Giordano d'Aloisio}.} \bibinfo{year}{2025}\natexlab{c}.
\newblock \bibinfo{title}{MANILA Video Demonstration}.
\newblock
\newblock
\urldef\tempurl%
\url{https://youtu.be/f2aXkk56DfY}
\showURL{%
\tempurl}


\bibitem[d'Aloisio et~al\mbox{.}(2023)]%
        {manila2023}
\bibfield{author}{\bibinfo{person}{Giordano d'Aloisio}, \bibinfo{person}{Antinisca {Di Marco}}, {and} \bibinfo{person}{Giovanni Stilo}.} \bibinfo{year}{2023}\natexlab{}.
\newblock \showarticletitle{Democratizing Quality-Based Machine Learning Development through Extended Feature Models}. In \bibinfo{booktitle}{\emph{Fundamental Approaches to Software Engineering}}, \bibfield{editor}{\bibinfo{person}{Leen Lambers} {and} \bibinfo{person}{Sebasti{\'a}n Uchitel}} (Eds.). Springer Nature Switzerland Cham, \bibinfo{publisher}{Springer Nature Switzerland}, \bibinfo{address}{Cham}, \bibinfo{pages}{88--110}.
\newblock
\showISBNx{978-3-031-30826-0}


\bibitem[d’Aloisio et~al\mbox{.}(2025)]%
        {d2024fair}
\bibfield{author}{\bibinfo{person}{Giordano d’Aloisio}, \bibinfo{person}{Claudio Di~Sipio}, \bibinfo{person}{Antinisca Di~Marco}, {and} \bibinfo{person}{Davide Di~Ruscio}.} \bibinfo{year}{2025}\natexlab{}.
\newblock \showarticletitle{How fair are we? From conceptualization to automated assessment of fairness definitions}.
\newblock \bibinfo{journal}{\emph{Software and Systems Modeling}} (\bibinfo{year}{2025}), \bibinfo{pages}{1--27}.
\newblock


\bibitem[d’Aloisio et~al\mbox{.}(2023)]%
        {daloisio_debiaser_2023}
\bibfield{author}{\bibinfo{person}{Giordano d’Aloisio}, \bibinfo{person}{Andrea D’Angelo}, \bibinfo{person}{Antinisca Di~Marco}, {et~al\mbox{.}}} \bibinfo{year}{2023}\natexlab{}.
\newblock \showarticletitle{Debiaser for {Multiple} {Variables} to enhance fairness in classification tasks}.
\newblock \bibinfo{journal}{\emph{Information Processing \& Management}} \bibinfo{volume}{60}, \bibinfo{number}{2} (\bibinfo{year}{2023}), \bibinfo{pages}{103226}.
\newblock
\showISSN{0306-4573}
\urldef\tempurl%
\url{https://doi.org/10.1016/j.ipm.2022.103226}
\showDOI{\tempurl}


\bibitem[Ferger(1931)]%
        {Ferger193136}
\bibfield{author}{\bibinfo{person}{Wirth~F. Ferger}.} \bibinfo{year}{1931}\natexlab{}.
\newblock \showarticletitle{The Nature and Use of the Harmonic Mean}.
\newblock \bibinfo{journal}{\emph{J. Amer. Statist. Assoc.}} \bibinfo{volume}{26}, \bibinfo{number}{173} (\bibinfo{year}{1931}), \bibinfo{pages}{36 – 40}.
\newblock
\urldef\tempurl%
\url{https://doi.org/10.1080/01621459.1931.10503148}
\showDOI{\tempurl}


\bibitem[Giagkiozis and Fleming(2014)]%
        {10.1162/EVCO_a_00128}
\bibfield{author}{\bibinfo{person}{Ioannis Giagkiozis} {and} \bibinfo{person}{Peter~J. Fleming}.} \bibinfo{year}{2014}\natexlab{}.
\newblock \showarticletitle{Pareto Front Estimation for Decision Making}.
\newblock \bibinfo{journal}{\emph{Evolutionary Computation}} \bibinfo{volume}{22}, \bibinfo{number}{4} (\bibinfo{year}{2014}), \bibinfo{pages}{651--678}.
\newblock
\showISSN{1063-6560}
\urldef\tempurl%
\url{https://doi.org/10.1162/EVCO_a_00128}
\showDOI{\tempurl}
\showeprint{https://direct.mit.edu/evco/article-pdf/22/4/651/1530439/evco\_a\_00128.pdf}


\bibitem[{Goncalves Jr.} and {Barros}(2011)]%
        {6086455}
\bibfield{author}{\bibinfo{person}{P.~M. {Goncalves Jr.}} {and} \bibinfo{person}{R.~S.~M. {Barros}}.} \bibinfo{year}{2011}\natexlab{}.
\newblock \showarticletitle{Automating Data Preprocessing with DMPML and KDDML}. In \bibinfo{booktitle}{\emph{2011 10th IEEE/ACIS International Conference on Computer and Information Science}}. \bibinfo{pages}{97--103}.
\newblock
\urldef\tempurl%
\url{https://doi.org/10.1109/ICIS.2011.23}
\showDOI{\tempurl}


\bibitem[Grossi et~al\mbox{.}(2018)]%
        {grossi_data_2018}
\bibfield{author}{\bibinfo{person}{Valerio Grossi}, \bibinfo{person}{Beatrice Rapisarda}, \bibinfo{person}{Fosca Giannotti}, {et~al\mbox{.}}} \bibinfo{year}{2018}\natexlab{}.
\newblock \showarticletitle{Data science at {SoBigData}: the {European} research infrastructure for social mining and big data analytics}.
\newblock \bibinfo{journal}{\emph{International Journal of Data Science and Analytics}} \bibinfo{volume}{6}, \bibinfo{number}{3} (\bibinfo{year}{2018}), \bibinfo{pages}{205--216}.
\newblock
\showISSN{2364-4168}
\urldef\tempurl%
\url{https://doi.org/10.1007/s41060-018-0126-x}
\showDOI{\tempurl}


\bibitem[He et~al\mbox{.}(2021)]%
        {HE2021106622}
\bibfield{author}{\bibinfo{person}{Xin He}, \bibinfo{person}{Kaiyong Zhao}, {and} \bibinfo{person}{Xiaowen Chu}.} \bibinfo{year}{2021}\natexlab{}.
\newblock \showarticletitle{AutoML: A survey of the state-of-the-art}.
\newblock \bibinfo{journal}{\emph{Knowledge-Based Systems}}  \bibinfo{volume}{212} (\bibinfo{year}{2021}), \bibinfo{pages}{106622}.
\newblock
\showISSN{0950-7051}
\urldef\tempurl%
\url{https://doi.org/10.1016/j.knosys.2020.106622}
\showDOI{\tempurl}


\bibitem[Johnson and Brun(2022)]%
        {johnson_fairkit-learn_2022}
\bibfield{author}{\bibinfo{person}{Brittany Johnson} {and} \bibinfo{person}{Yuriy Brun}.} \bibinfo{year}{2022}\natexlab{}.
\newblock \showarticletitle{Fairkit-learn: a fairness evaluation and comparison toolkit}. In \bibinfo{booktitle}{\emph{Proceedings of the {ACM}/{IEEE} 44th {International} {Conference} on {Software} {Engineering}: {Companion} {Proceedings}}} \emph{(\bibinfo{series}{{ICSE} '22})}. \bibinfo{publisher}{Association for Computing Machinery}, \bibinfo{address}{New York, NY, USA}, \bibinfo{pages}{70--74}.
\newblock
\showISBNx{978-1-4503-9223-5}
\urldef\tempurl%
\url{https://doi.org/10.1145/3510454.3516830}
\showDOI{\tempurl}


\bibitem[Lee and Singh(2021)]%
        {lee_landscape_2021}
\bibfield{author}{\bibinfo{person}{Michelle Seng~Ah Lee} {and} \bibinfo{person}{Jat Singh}.} \bibinfo{year}{2021}\natexlab{}.
\newblock \showarticletitle{The {Landscape} and {Gaps} in {Open} {Source} {Fairness} {Toolkits}}. In \bibinfo{booktitle}{\emph{Proceedings of the 2021 {CHI} {Conference} on {Human} {Factors} in {Computing} {Systems}}} \emph{(\bibinfo{series}{{CHI} '21})}. \bibinfo{publisher}{Association for Computing Machinery}, \bibinfo{address}{New York, NY, USA}, \bibinfo{pages}{1--13}.
\newblock
\showISBNx{978-1-4503-8096-6}
\urldef\tempurl%
\url{https://doi.org/10.1145/3411764.3445261}
\showDOI{\tempurl}


\bibitem[Lerman(1980)]%
        {lerman1980fitting}
\bibfield{author}{\bibinfo{person}{PM Lerman}.} \bibinfo{year}{1980}\natexlab{}.
\newblock \showarticletitle{Fitting segmented regression models by grid search}.
\newblock \bibinfo{journal}{\emph{Journal of the Royal Statistical Society Series C: Applied Statistics}} \bibinfo{volume}{29}, \bibinfo{number}{1} (\bibinfo{year}{1980}), \bibinfo{pages}{77--84}.
\newblock


\bibitem[Liu et~al\mbox{.}(2015)]%
        {liu2015survey}
\bibfield{author}{\bibinfo{person}{Ji Liu}, \bibinfo{person}{Esther Pacitti}, \bibinfo{person}{Patrick Valduriez}, {et~al\mbox{.}}} \bibinfo{year}{2015}\natexlab{}.
\newblock \showarticletitle{A survey of data-intensive scientific workflow management}.
\newblock \bibinfo{journal}{\emph{Journal of Grid Computing}} \bibinfo{volume}{13}, \bibinfo{number}{4} (\bibinfo{year}{2015}), \bibinfo{pages}{457--493}.
\newblock


\bibitem[Martínez-Fernández et~al\mbox{.}(2022)]%
        {martinez-fernandez_software_2022}
\bibfield{author}{\bibinfo{person}{Silverio Martínez-Fernández}, \bibinfo{person}{Justus Bogner}, \bibinfo{person}{Xavier Franch}, {et~al\mbox{.}}} \bibinfo{year}{2022}\natexlab{}.
\newblock \showarticletitle{Software {Engineering} for {AI}-{Based} {Systems}: {A} {Survey}}.
\newblock \bibinfo{journal}{\emph{ACM Transactions on Software Engineering and Methodology}} \bibinfo{volume}{31}, \bibinfo{number}{2} (\bibinfo{year}{2022}), \bibinfo{pages}{37e:1--37e:59}.
\newblock
\showISSN{1049-331X}
\urldef\tempurl%
\url{https://doi.org/10.1145/3487043}
\showDOI{\tempurl}


\bibitem[Mehrabi et~al\mbox{.}(2021)]%
        {mehrabi_survey_2021}
\bibfield{author}{\bibinfo{person}{Ninareh Mehrabi}, \bibinfo{person}{Fred Morstatter}, \bibinfo{person}{Nripsuta Saxena}, {et~al\mbox{.}}} \bibinfo{year}{2021}\natexlab{}.
\newblock \showarticletitle{A Survey on Bias and Fairness in Machine Learning}.
\newblock \bibinfo{journal}{\emph{ACM Comput. Surv.}} \bibinfo{volume}{54}, \bibinfo{number}{6}, Article \bibinfo{articleno}{115} (\bibinfo{year}{2021}), \bibinfo{numpages}{35}~pages.
\newblock
\showISSN{0360-0300}
\urldef\tempurl%
\url{https://doi.org/10.1145/3457607}
\showDOI{\tempurl}


\bibitem[Muccini and Vaidhyanathan(2021)]%
        {muccini_software_2021}
\bibfield{author}{\bibinfo{person}{Henry Muccini} {and} \bibinfo{person}{Karthik Vaidhyanathan}.} \bibinfo{year}{2021}\natexlab{}.
\newblock \showarticletitle{Software {Architecture} for {ML}-based {Systems}: {What} {Exists} and {What} {Lies} {Ahead}}. In \bibinfo{booktitle}{\emph{2021 {IEEE}/{ACM} 1st {Workshop} on {AI} {Engineering} - {Software} {Engineering} for {AI} ({WAIN})}}. \bibinfo{pages}{121--128}.
\newblock
\urldef\tempurl%
\url{https://doi.org/10.1109/wain52551.2021.00026}
\showDOI{\tempurl}


\bibitem[Nguyen et~al\mbox{.}(2024)]%
        {nguyen_literature_2024}
\bibfield{author}{\bibinfo{person}{Thanh Nguyen}, \bibinfo{person}{Maria~Teresa Baldassarre}, \bibinfo{person}{Luiz~Fernando de Lima}, {and} \bibinfo{person}{Ronnie de Souza~Santos}.} \bibinfo{year}{2024}\natexlab{}.
\newblock \showarticletitle{From {Literature} to {Practice}: {Exploring} {Fairness} {Testing} {Tools} for the {Software} {Industry} {Adoption}}. In \bibinfo{booktitle}{\emph{Proceedings of the 18th {ACM}/{IEEE} {International} {Symposium} on {Empirical} {Software} {Engineering} and {Measurement}}} \emph{(\bibinfo{series}{{ESEM} '24})}. \bibinfo{publisher}{Association for Computing Machinery}, \bibinfo{address}{New York, NY, USA}, \bibinfo{pages}{549--555}.
\newblock
\showISBNx{9798400710476}
\urldef\tempurl%
\url{https://doi.org/10.1145/3674805.3695404}
\showDOI{\tempurl}


\bibitem[{ONU}({[n.\,d.]})]%
        {onu_sdg}
\bibfield{author}{\bibinfo{person}{{ONU}}.} \bibinfo{year}{[n.\,d.]}\natexlab{}.
\newblock \bibinfo{title}{{ONU Sustainable Development Goals}}.
\newblock
\newblock
\urldef\tempurl%
\url{https://www.un.org/sustainabledevelopment/}
\showURL{%
\tempurl}


\bibitem[PalletsProject(2023)]%
        {jinja2}
\bibfield{author}{\bibinfo{person}{PalletsProject}.} \bibinfo{year}{2023}\natexlab{}.
\newblock \bibinfo{title}{Jinja website}.
\newblock
\newblock
\urldef\tempurl%
\url{https://jinja.palletsprojects.com/}
\showURL{%
\tempurl}


\bibitem[Pedregosa et~al\mbox{.}(2011)]%
        {scikit-learn}
\bibfield{author}{\bibinfo{person}{F. Pedregosa}, \bibinfo{person}{G. Varoquaux}, \bibinfo{person}{A. Gramfort}, {et~al\mbox{.}}} \bibinfo{year}{2011}\natexlab{}.
\newblock \showarticletitle{Scikit-learn: Machine Learning in {P}ython}.
\newblock \bibinfo{journal}{\emph{Journal of Machine Learning Research}}  \bibinfo{volume}{12} (\bibinfo{year}{2011}), \bibinfo{pages}{2825--2830}.
\newblock


\bibitem[Runeson et~al\mbox{.}(2012)]%
        {runeson2012case}
\bibfield{author}{\bibinfo{person}{Per Runeson}, \bibinfo{person}{Martin Host}, \bibinfo{person}{Austen Rainer}, {et~al\mbox{.}}} \bibinfo{year}{2012}\natexlab{}.
\newblock \bibinfo{booktitle}{\emph{Case study research in software engineering: Guidelines and examples}}.
\newblock \bibinfo{publisher}{John Wiley \& Sons}.
\newblock


\bibitem[Rönkkö et~al\mbox{.}(2015)]%
        {RONKKO201513}
\bibfield{author}{\bibinfo{person}{Mauno Rönkkö}, \bibinfo{person}{Jani Heikkinen}, \bibinfo{person}{Ville Kotovirta}, {et~al\mbox{.}}} \bibinfo{year}{2015}\natexlab{}.
\newblock \showarticletitle{Automated preprocessing of environmental data}.
\newblock \bibinfo{journal}{\emph{Future Generation Computer Systems}}  \bibinfo{volume}{45} (\bibinfo{year}{2015}), \bibinfo{pages}{13--24}.
\newblock
\showISSN{0167-739X}
\urldef\tempurl%
\url{https://doi.org/10.1016/j.future.2014.10.011}
\showDOI{\tempurl}


\bibitem[Saleiro et~al\mbox{.}(2018)]%
        {saleiro2018aequitas}
\bibfield{author}{\bibinfo{person}{Pedro Saleiro}, \bibinfo{person}{Benedict Kuester}, \bibinfo{person}{Loren Hinkson}, {et~al\mbox{.}}} \bibinfo{year}{2018}\natexlab{}.
\newblock \showarticletitle{Aequitas: A bias and fairness audit toolkit}.
\newblock \bibinfo{journal}{\emph{arXiv preprint arXiv:1811.05577}} (\bibinfo{year}{2018}).
\newblock
\urldef\tempurl%
\url{http://arxiv.org/abs/1811.05577}
\showURL{%
\tempurl}


\bibitem[Wexler et~al\mbox{.}(2019)]%
        {wexler2019if}
\bibfield{author}{\bibinfo{person}{James Wexler}, \bibinfo{person}{Mahima Pushkarna}, \bibinfo{person}{Tolga Bolukbasi}, \bibinfo{person}{Martin Wattenberg}, \bibinfo{person}{Fernanda Vi{\'e}gas}, {and} \bibinfo{person}{Jimbo Wilson}.} \bibinfo{year}{2019}\natexlab{}.
\newblock \showarticletitle{The what-if tool: Interactive probing of machine learning models}.
\newblock \bibinfo{journal}{\emph{IEEE transactions on visualization and computer graphics}} \bibinfo{volume}{26}, \bibinfo{number}{1} (\bibinfo{year}{2019}), \bibinfo{pages}{56--65}.
\newblock


\bibitem[Yohannis and Kolovos(2022)]%
        {10.1145/3550355.3552401}
\bibfield{author}{\bibinfo{person}{Alfa Yohannis} {and} \bibinfo{person}{Dimitris Kolovos}.} \bibinfo{year}{2022}\natexlab{}.
\newblock \showarticletitle{Towards Model-Based Bias Mitigation in Machine Learning}. In \bibinfo{booktitle}{\emph{Proceedings of the 25th International Conference on Model Driven Engineering Languages and Systems}} (Montreal, Quebec, Canada) \emph{(\bibinfo{series}{Models '22})}. \bibinfo{publisher}{Association for Computing Machinery}, \bibinfo{address}{New York, NY, USA}, \bibinfo{pages}{143–153}.
\newblock
\showISBNx{9781450394666}
\urldef\tempurl%
\url{https://doi.org/10.1145/3550355.3552401}
\showDOI{\tempurl}


\end{thebibliography}
\end{document}